\documentclass[10pt,letterpaper,twocolumn]{article} 

\usepackage{ol2}
\usepackage[draft]{hyperref}
\usepackage{amsmath}

\begin{document}

\twocolumn[ 

\title{Fiber amplification of radially and azimuthally polarized laser light}


\author{Moti Fridman$^1$, Micha Nixon$^1$, Mark Dubinskiy$^2$, Asher A. Friesem$^1$, and Nir Davidson$^1$}

\address{$^1$ Weizmann Institute of Science, Dept. of Physics of Complex Systems, Rehovot 76100, Israel
$^*$Corresponding author: moti.fridman@weizmann.ac.il }

\address{$^2$ US Army Research Lab Adelphi, MD 20783, USA}

\begin{abstract}
The results on amplifying either radially or azimuthally polarized
light with a fiber amplifier are presented. Experimental results
reveal that more than 85\% polarization purity can be retained at
the output even with 40dB amplification, and that efficient
conversion of the amplified light to linear polarization can be
obtained.
\end{abstract}


 ] 

\noindent

Radially and azimuthally polarized laser light have unique
properties and symmetries that are advantageous for high power
lasers~\cite{GalinaSVRHigh, GalinaAmp}. For example, their cross
section intensity distributions have doughnut shapes whose peak
intensities are significantly lower than that of light with a
Gaussian distribution, so non-linear and damage effects through
fibers core are significantly reduced~\cite{MyRadial}.
Accordingly, several laser resonator configurations have been
developed to produce radially and azimuthally polarized
light~\cite{Oron, Grosjean2, ComputerHologram,
ShirakwaRadialFiber1, ShirakwaRadialFiber2, GalinaSVR, SVR2008,
MyRadial}. However, in all these the power of the output beam was
relatively low, so that for some applications additional
amplification is required~\cite{LaserProssessing, Drill}. Thus
far, amplification of either radially or azimuthally polarized
laser light was demonstrated with bulk solid state laser
amplifiers~\cite{GalinaAmp}, but not with fiber amplifiers,
probably because of deleterious effects of their birefringence and
non-linearities. Here we present for the first time experimental
results on amplifying radially and azimuthally polarized light
with a large mode area fiber amplifier, where we obtained an
amplification of 40dB with over $85\%$ polarization purity. We
also show that the amplified beam can be efficiently converted
back into a linearly polarized near-Gaussian beam, confirming the
modal structure and polarization purity of the radially and
azimuthally polarized amplified beam.

\begin{figure}[htb]
\centerline{\includegraphics[width=8cm]{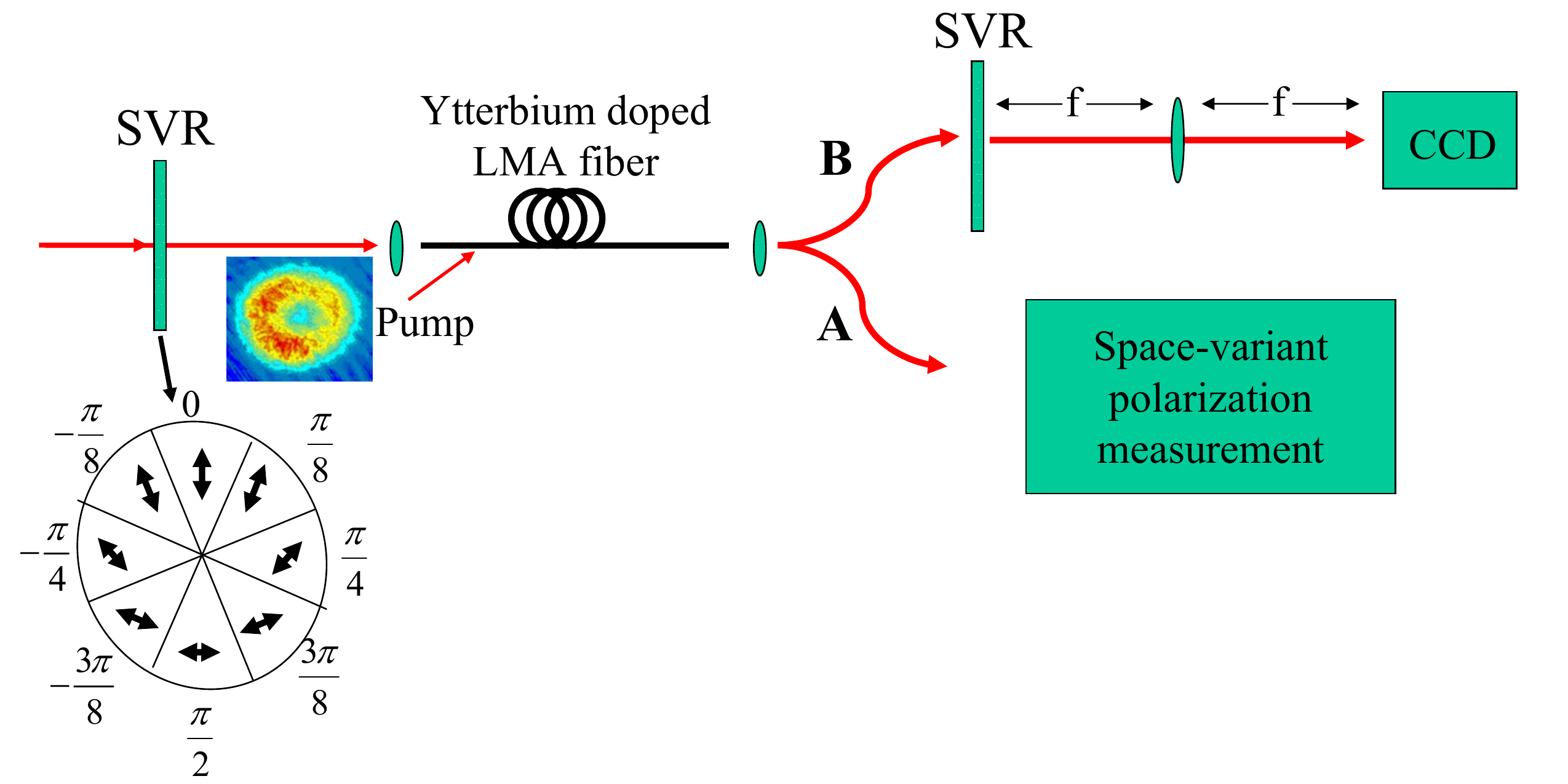}}
\caption{\label{setup}Experimental configuration for amplifying
radially and azimuthally polarized light with a fiber amplifier.
Path A - space variant polarization measurement, path B - conversion
arrangement to a linearly polarized light.}
\end{figure}

The experimental arrangement for amplifying either radially or
azimuthally polarized beam with a fiber amplifier is presented
schematically in Fig.~\ref{setup}, along with an arrangement for
converting the amplified radial or azimuthal light distribution to
a Gaussian distribution. In this configuration a weak linearly
polarized Gaussian beam of $40\mu W$ propagated through a
space-variant retardation plate(SVR) to obtain either radially or
azimuthally polarized light that served as the
input~\cite{GalinaSVR, SVR2008, MyRadial}. The SVR was comprised
of eight sectors each having $\lambda / 2$ retardation in
different orientations, where the direction of the slow axis of
each retardation sector is denoted by an arrow as shown in
Fig.~\ref{setup}. When the SVR was oriented at $0^{\circ}$ with
respect to the polarization of the input beam, it converted the
linearly polarized light into radially polarized light, and when
the SVR was oriented at $45^{\circ}$ it converted the linearly
polarized light to azimuthally polarized light. In both cases the
spatial mode of the beam was converted into a $LG_{10}^{*}$ mode.

The radially or azimuthally polarized beam was then lunched into a
$10m$ long Ytterbium doped double clad fiber, with core diameter
of $20 \mu m$ and numerical aperture of $0.07$, that served as the
amplifier. Such a fiber amplifier can support the propagation of
four modes, and in particular the $TE_{01}$ and $TM_{01}$ modes of
the radial and azimuthal polarizations inside the fiber, which
corresponds to the $LG_{10}^{*}$ mode in free
space~\cite{AgrawalModes, Volpe}. The fiber amplifier was
co-pumped with a $915nm$ diode laser.

\begin{figure}[htb]
\centerline{\includegraphics[width=8cm]{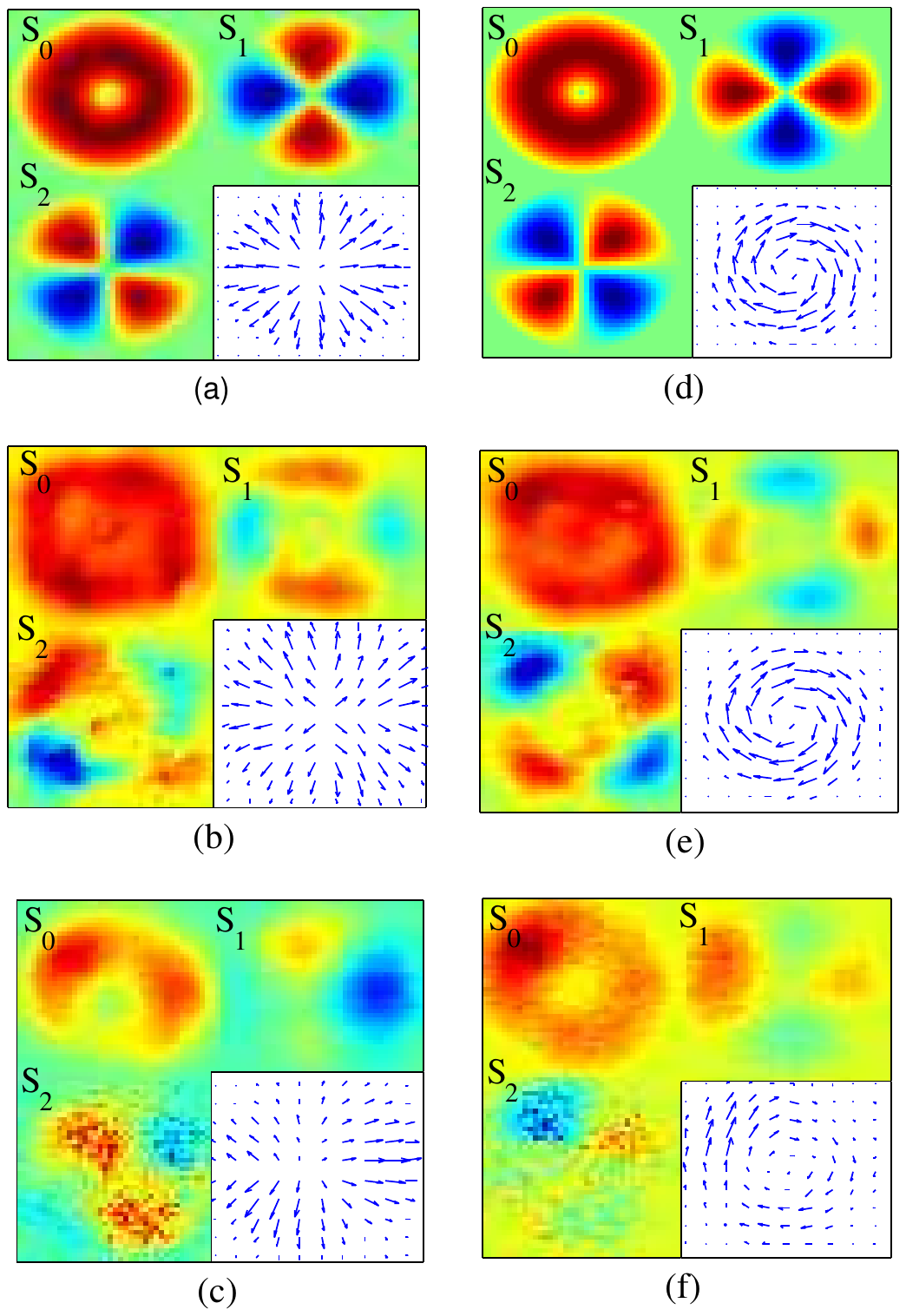}}
\caption{\label{all}Calculated and experimental Stokes parameters
and the state of polarization of input and amplified radially and
azimuthally polarized light beams. (a) Calculated radially polarized
light; (b) experimental radially polarized input beam; (c)
experimental radially polarized amplified beam; (d) calculated
azimuthally polarized beam; (e) experimental azimuthally polarized
input beam; (f) experimental azimuthally polarized amplified beam.
The arrows represent the resulting main axis of the local
polarization ellipse.}
\end{figure}

We measured the amplified output power and found it to be 400mW,
indicating that the amplification was 40dB. The polarization
distributions of the input and amplified light beams were
determined by measuring their intensity distributions after
passing through three different sets of polarization elements, to
obtain the four Stokes parameters~\cite{Hecht}. Some
representative results of three of the four Stokes parameters
together with the resulting polarization state at each point in
the beams, are shown in Fig.~\ref{all}. Figures~\ref{all}(a), (b)
and (c) show the three Stokes parameters and the state of
polarization for the calculated, input and output light
distributions with radial polarization. Figures~\ref{all}(d), (e)
and (f) show the three Stokes parameters and the state of
polarization for the calculated, input and output light
distributions with azimuthal polarization. As evident, there is
good agreement between the calculated results and the experimental
results for the input and amplified beams.

Using the Stokes parameters, we calculated the polarization purity
of the experimentally detected beams and found that for both the
radially and azimuthally polarized input beams the polarization
purity was over $95\%$, and for the corresponding amplified beams
the polarization purity was over $85\%$. These results obviously
indicate that polarization purity is essentially preserved. We
found that the amplified radially and azimuthally polarized light
beams were essentially stable for several hours in terms of
polarization purity and intensity distributions. The experiments
were conducted in usual laboratory environment, with minimal
mechanical stresses on the fiber amplifier.

\begin{figure}[htb]
\centerline{\includegraphics[width=8cm]{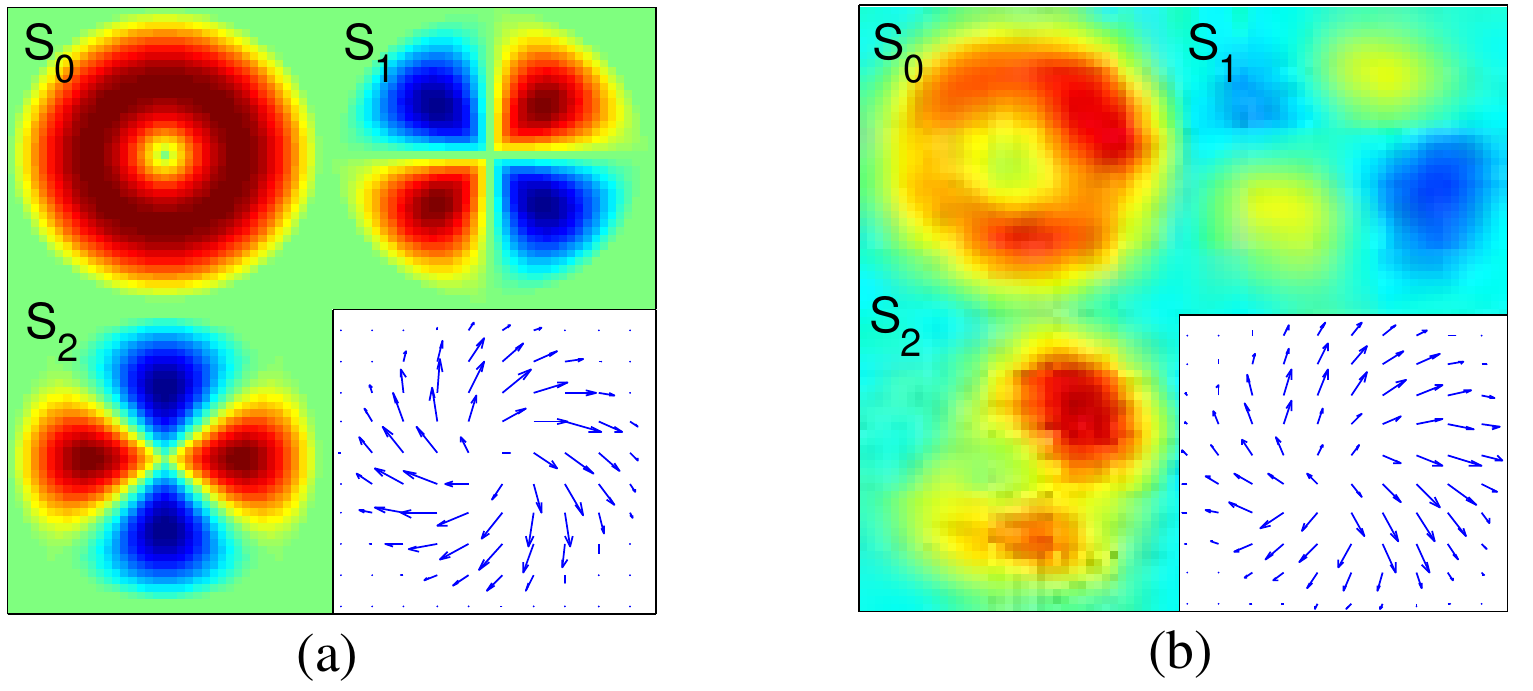}}
\caption{\label{comb}Calculated and experimental Stokes parameters
and the state of polarization of amplified combination of radially
and azimuthally polarized beam. (a) Calculated results; (b)
experimental results.}
\end{figure}

We also amplified a linear combination of radially and azimuthally
polarized input light beam (generated when the SVR is oriented at
$22.5^{\circ}$), and detected the Stokes parameters and the
corresponding full state of polarization of the amplified beam.
The results are presented in Fig.~\ref{comb}. Figure~\ref{comb}(a)
shows the calculated results while Fig.~\ref{comb}(b) shows the
experimental results. Here again there is good agreement between
the calculated and experimental results. The corresponding
polarization purity for the experimentally amplified output beam
was calculated to be over $85\%$.

\begin{figure}[htb]
\centerline{\includegraphics[width=8cm]{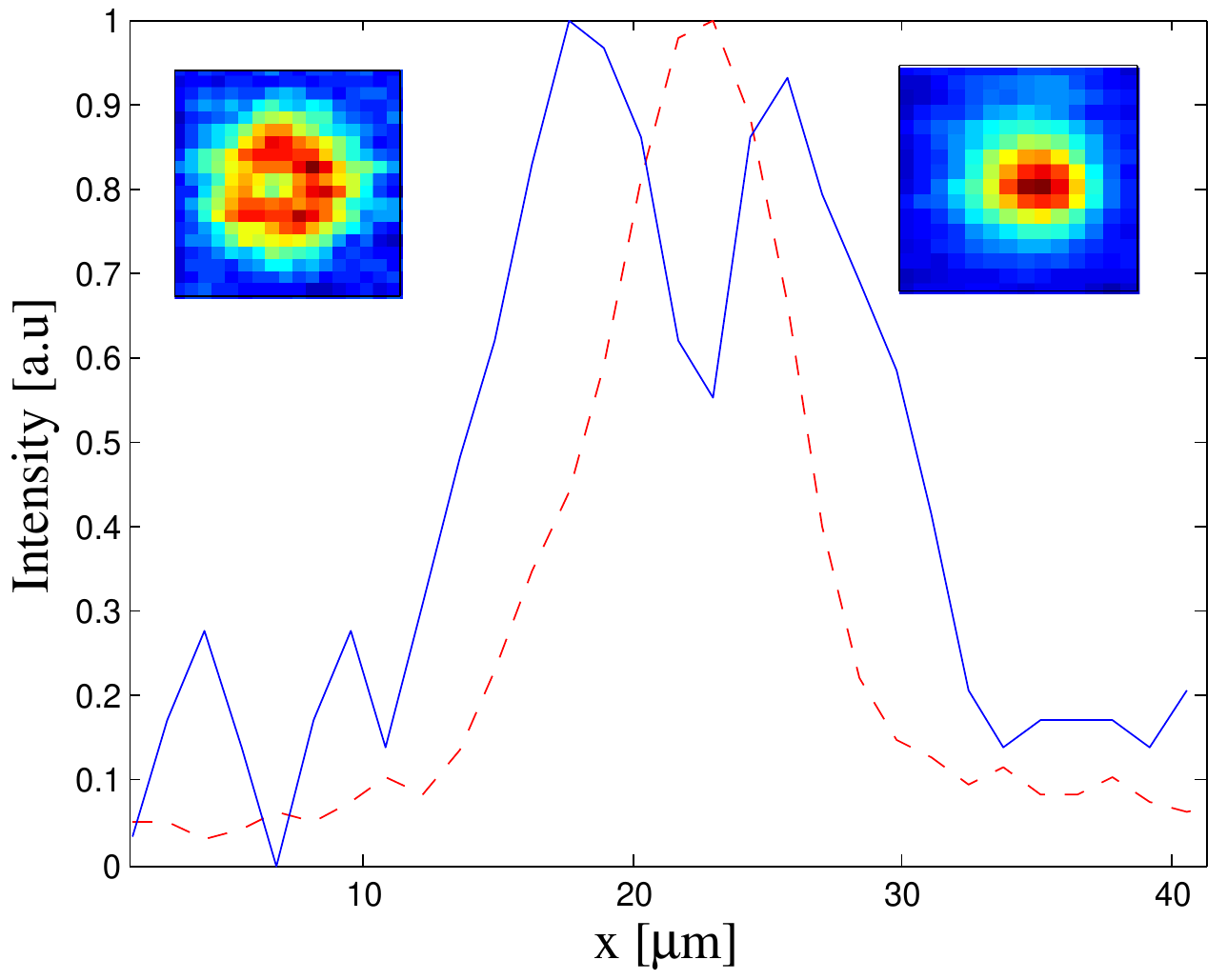}}
\caption{\label{FF}Experimental cross-sections of the far-field
intensity distributions of the amplified radially polarized light
beam with and without conversion to a linearly polarized near
Gaussian beam. Solid (blue) curve - before conversion; dashed (red)
curve - after conversion. Left inset depicts the far field intensity
distribution before conversion and right inset the far field
intensity distribution after conversion.}
\end{figure}

Finally, we converted the amplified radially polarized light beam
into a linearly polarized light beam with Gaussian distribution by
means of an additional SVR, and detected the far-field intensity
distribution by means of a focusing lens and a CCD camera as shown
in path B in Fig.~\ref{setup}. The results are presented in
Fig.~\ref{FF}. It shows the experimental cross-sections of the
amplified radially polarized light beam in the far-field, with and
without conversion. Also shown are the actual far-field intensity
distributions of the amplified radially polarized light without
conversion (upper left inset) and that of the linearly polarized
light with conversion (upper right inset). As evident, the far field
intensity distribution of the radially polarized beam has the
expected doughnut shape, reminiscent of the $LG_{10}^{*}$ mode
distribution, and that of the converted linearly polarized beam has
the expected Gaussian distribution. The conversion to a Gaussian or
near Gaussian distribution with measured $85\%$ polarization purity
indicates that the amplified radially polarized light beam was
indeed of high purity.

To conclude, we presented a novel configuration for amplifying
radially and azimuthally polarized light, obtaining 40dB
amplification while maintaining the polarization purity at over
$85\%$. We also showed that the amplified radially polarized light
beam can be efficiently converted to a linearly polarized light beam
with near Gaussian distribution. All our experiments were done with
relatively low power levels, where thermally induced stresses are
not significant. At higher power levels these will have to be taken
into account.

We thanks Galina Machavariani and Steve Jackel from Soreq Nuclear
Research Center for supplying the SVR elements.


\end{document}